\documentclass[aps,pre,twocolumn,showpacs,floatfix,preprintnumbers,amsmath,amssymb]{revtex4}
\usepackage{amsmath}
\usepackage{amssymb}
\usepackage{epsfig}
\usepackage{graphicx}
\usepackage{dcolumn}
\usepackage{bm}

\begin{document}
\title{Dynamics of kicked particles in a double-barrier structure}
\author{Harinder Pal$^a$\footnote{harinder@prl.res.in} and M. S. Santhanam$^b$\footnote{santh@iiserpune.ac.in}}
\affiliation{$^a$Physical Research Laboratory, Navrangpura, Ahmedabad 380 009, India.\\
$^b$Indian Institute of Science Education and Research, Pashan Road, Pune 411 021, India.}

\date{\today}
\begin{abstract}
We study the classical and quantum dynamics of periodically kicked particles placed
initially within an open, double barrier structure. This system does not obey
the Kolmogorov-Arnold-Moser (KAM) theorem and displays chaotic dynamics.
The phase space features induced by non-KAM nature of the system leads to 
dynamical features such as the non-equilibrium steady
state, classically induced saturation of energy growth and momentum filtering.
We also comment on the experimental feasibility of this system as well as its
relevance in the context of current interest in classically induced
localisation and chaotic ratchets.

\end{abstract}

\maketitle

\section{introduction}
Periodically kicked rotor is a popular model that has served
as a paradigm to understand Hamiltonian chaos both in the
classical and quantum regime \cite{chi1}. This was originally introduced
as a simple model for dynamical chaos but was sufficiently
general enough to cover many physical situations. For instance,
problems like the Hydrogen atom in microwave fields and motion of
a comet around the sun driven by a suitable planet
can be reduced to that of kicked rotor \cite{chi2}. This system
is also paradigmatic for another important reason; it obeys KAM (Kolmogorov-
Arnold-Moser) theorem \cite{kam1}. This implies that, as a control parameter
is varied, the transition from regularity to chaos occurs progressively by
breaking of invariant curves in phase space. Once all the invariant curves
are broken down, diffusive global transport of particles in phase space
becomes possible. In the corresponding quantum regime, this classical
diffusive transport is inhibited by the onset of dynamical localisation \cite{cas1}.
This was experimentally realised in the laboratory with cold atoms in
optical lattices \cite{mark1} and is the basis for theoretical and
experimental realization of chaotic ratchets in recent times \cite{rat}.

On the other hand, there are other physical systems that exhibit classical
chaos but violate the KAM theorem, the so-called non-KAM systems.
This class includes the kicked harmonic oscillator (KHO) \cite{zas} and the kicked particle in
infinite square well potential \cite{sankar, bhu}. In both these cases,
when a parameter is varied, the invariant curves
are replaced by stochastic webs \cite{zas}, an intricate chain of islands and
globally connected channels, through which particle transport becomes possible.
Non-KAM type is also relevant for an important class of physical systems, namely
the dynamics of particles in quantum wells and barrier structures.
Till date, non-KAM systems have been
experimentally realised in semiconductor superlattices in tilted
magnetic fields in which the enhanced
conductivity could be attributed to non-KAM chaos \cite{from}.
Further, measurement of Lochschmidt echo using a non-KAM system, namely,
the ion trap with harmonic
potential in the presence of a kicking field \cite{gar} has also been proposed.
Inspite of this,
very few non-KAM systems have been investigated and they have not been explored
in sufficient details.

Another motivation for this work stems from the considerable interest in
recent times in the dynamics of condensates
placed in finite box-type potentials acted upon by a periodically kicking field.
In a recent experiment, Henderson et. al. \cite{mark2} have constructed a quasi-1D finite box,
using a combination of optical and magnetic trap, with the Bose-Einstein condensates
in the box receiving periodic kicks. This set-up was used to study
the effect of atomic interactions on the transport of BECs. In place of the
dynamical localisation they observed a classical saturation in the energy of BECs
due to a balance between the energy gained from kicks and the energy lost by
leakage of BECs over the finite barrier \cite{mark2}.
Apart from this, a series of experiments \cite{bec-trans} that studied the transport of BECs
in the presence of disordered potential have reported such classically induced
energy saturation effects.
Then the question is if it is possible to observe such classically
induced energy saturation in chaotic systems without inter-particle interactions and
what would its mechanism be ?
We show that kicked particle in finite well type potential that we study in this paper
shows this feature and we discuss its mechanism.
It is also relevant to point out that following the achievement of
BECs in the optical box trap \cite{mark3}, theoretical investigations of resonance and
anti-resonance behaviour and its relation to the KAM and non-KAM type
dynamics for BECs in 1D infinite well have also been performed \cite{bli}.
Further, experiments exploring the interface of nonlinear dynamics of electrons
in 1D quantum well irradiated at terahertz frequencies have already been
reported \cite{sherwin}.

Although non-KAM type dynamics is a generic feature in physical systems such
as the potential wells not much work has been done on this class of problems.
However, on the theoretical front, infinite square well potential confining a
delta-kicked particle has been studied \cite{sankar,bhu}.
In the light of recent attempts to study chaotic ratchets \cite{rat}, in both the classical
and quantum sense, it would be useful to open the potential to allow for
particle transport. This could lead to chaotic ratchets that can utilise
non-KAM features of its classical system for directed transport.
In this paper, we study a periodically kicked particle initially held
in between a finite double barrier structure. Double barrier heterostructures
play an important role in electronic devices that use resonant tunnelling diodes \cite{phar}
though without the kicking potential.
Primarily we present numerical explorations of this
problem to study its rich dynamical features. In the next section, we introduce our model and
in subsequent sections we discuss the classical and quantum dynamics of this system.

\section{The kicked particle in double barrier}
\begin{figure}
\includegraphics*[width=2.8in]{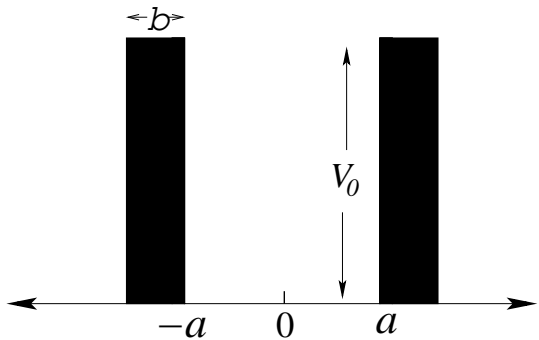}
\caption{Schematic of the stationary part of the potential. The width of each barrier is $b$.
The well region has width $2a$.}
\label{scheme}
\end{figure}
We consider the dynamics of a non-interacting particle initially located in between
two potential barriers each of height $V_0$ and width $b$ distant
$2a$ apart (see Fig. \ref{scheme}). The particle is further subjected to
flashing $\delta$-kicks of period $T$ generated by a spatially periodic
potential field of wavelength $\lambda$.
Amplitude $\epsilon$ of the kicking field is generally referred to as kick strength.
The classical Hamiltonian of the system is
\begin{equation}
\widetilde{H} = \frac{\widetilde{p}^2}{2m} + \widetilde{V}_{sq} + \widetilde{\epsilon} \cos\left( \frac{2\pi \widetilde{x}}{\lambda} +
\phi \right) \sum_{n=-\infty}^{\infty} \delta(\widetilde{t}-nT),
\label{ham}
\end{equation}
where
$\widetilde{V}_{sq} = \widetilde{V}_0 \left[ \Theta(\widetilde{x}+\widetilde{b}+a) -
\Theta(\widetilde{x}+a)+ \Theta(\widetilde{x}-a)\right.$ 
$\left.-\Theta(\widetilde{x}-a-\widetilde{b}) \right]$,
$\Theta(.)$ is the unit step function and $\phi$ is the phase of the kicking field.
The set of canonical transformations given by,
\begin{eqnarray}
\widetilde{t}=tT, \;\;\;\; \widetilde{x}= \lambda \frac{(x-\phi)}{2\pi},
\;\;\;\; \widetilde{p}=\frac{p T E_c}{\lambda \pi}, \nonumber \\
\widetilde{H}=\frac{H E_c}{2\pi^2},
\;\;\;\; \widetilde{\epsilon}=\frac{\epsilon E_c T}{2\pi^2},
\;\;\;\; \widetilde{V}_0=\frac{V_0 E_c}{2\pi^2},
\;\;\;\; \widetilde{b} = \frac{\lambda}{2\pi} b,
\label{transform}
\end{eqnarray}
with $E_c = m \lambda^2/2 T^2$
leads to a new dimensionless Hamiltonian
\begin{equation}
H = \frac{{p}^2}{2} + V_{sq} + \epsilon \cos\left(x \right)
\sum_{n=-\infty}^{\infty} \delta(t-n).
\label{sham}
\end{equation}
In this, ${V_{sq}} ={V_0} \left[ \Theta(x-\phi+b+R\pi)-\Theta(x-\phi+R\pi)\right.$ 
$\left.+\Theta(x-\phi-R\pi)-\Theta(x-\phi -R\pi-b) \right]$ with $R=2a/\lambda$ being the ratio of
the distance between the barriers to the wave length of the kicking field. The classical dynamics of the
system depends upon five parameters, namely, $\epsilon, R, b, {V_0}$ and $\phi$.
Of these, $R,b$ and $\phi$ determine the positions of discontinuities in the potential (position of
the wall boundaries) at $\textbf{B}=\{-x_l-b, -x_l, x_r, x_r+b \}$ where $x_l = -R \pi + \phi$ and $x_r = R \pi + \phi$.
Note that if $\phi=0$, then $x_l = -x_w$ and $x_r = x_w$ with $x_w = R \pi$.
Thus, the qualitative nature of the classical dynamics depends on the positions of
wall boundaries collectively denoted by $\textbf{B}$, the kick strength $\epsilon$ and potential height $V_0$.
In this paper (except in section III(C)), we set $\phi=0$ which makes the potential symmetric about $x=0$.
It is useful to write Eq. \ref{sham} as
\begin{equation}
H=H_0+V_{sq}(x),
\end{equation}
where $H_0=\frac{{p}^2}{2} + \epsilon \cos\left(x \right)
\sum_{n=-\infty}^{\infty} \delta(t-n)$
leads to standard map defined on the infinite plane. Note that if $V_{sq}(x)=V_0$, a constant, then
the Hamilton's equations will not have the potential term and the dynamics would be
completely governed by $H_0$.

\section{Classical Dynamics}
\subsection{Classical map}
The Hamiltonian in Eq. (\ref{sham}) is classically integrable for $\epsilon=0$.
This corresponds to free motion in the presence of two potential barriers and hence it
is possible to obtain a transformation to action-angle variables.
For $\epsilon > 0$, the system is non-integrable and can even display abrupt transition to
chaotic dynamics with mixed phase space depending on the values of $R,b$ and $\phi$.
It is convenient to think of the system as being entirely governed by $H_0$
and then incorporate effect of discontinuities in $V_{sq}$ through appropriate
boundary conditions. This leads to the following map,
\begin{subequations}
\begin{eqnarray}
p_{n} & = & p_{n-1} + \epsilon \sin(x_{n-1}), \nonumber \\
x_{n} & = & x_{n-1} + p_{n},
\label{smap}
\end{eqnarray}
\begin{equation}
\left( \begin{array}{c} p_{n}\\x_{n} \end{array}\right) \to
\widehat{\mathcal{R}}\left( \begin{array}{c} p_{n} \\ x_{n} \end{array}\right).
\label{bc1}
\end{equation}
\label{smap1}
\end{subequations}
Equation \ref{smap} represents the effect of $H_0$ and is identical to the standard
map. In Eq. \ref{bc1}, the operator $\widehat{\mathcal{R}} = \widehat{\mathcal{R}}_k\ldots
\widehat{\mathcal{R}}_2 \widehat{\mathcal{R}}_1$ represents the effect due to
$k$ encounters of the particle, in between two kicks, with the discontinuities of $V_{sq}$ at positions 
represented by $B_1, B_2, \ldots B_k $ respectively.
Depending on the energy, each of these $k$ encounters could either be a reflection (sign of momentum
changes) or refraction (magnitude of momentum changes) at $B_i \in \textbf{B}$, $i=1,2,....k$.

The map in Eq. \ref{smap1} would be complete if the operator $\widehat{\mathcal{R}}_i$, 
that incorporates effect of $i^{th}$ discontinuity encountered, is explicitly written down.
Between successive kicks applied at times $n$ and $n+1$, 
we denote the state of the particle after incorporating effect of $i$th encounter
with a boundary $B_i$ by $\left(\begin{array}{c} x_{n}^{i}\\ p_{n}^{i} \end{array}\right)$.
We define $\left]x_s^{i}, x_n^{i} \right[$ with $i=0,1 \ldots k$ as the path, starting from $x_s^i$, a particle would 
traverse between the two kicks after encountering  
$i^{th}$ discontinuity if there were no discontinuities to be faced till the next kick.
For $i=0$, $x_s^{i}$ would simply be
$x_{n-1}$ and would be equal to $B_i$ for $i>0$. $x_n^0$ and $p_n^0$ to be used in boundary conditions would
simply be $x_n$ and $p_n$ obtained directly from Eq. \ref{smap}. Boundary conditions defined by Eq. 
\ref{bound} below are applied $k$ times until $\left]x_s^{k}, x_n^{k} \right[\cap \textbf{B}=\emptyset  $. If $E_n$ denotes
the energy of the system at $n$th kick, then for $E_n \leq V_0$
(reflective boundary condition), we obtain
\begin{subequations}
\begin{equation}
\widehat{\mathcal{R}}_i \left( \begin{array}{c} x_{n}^{i-1}\\\\p_{n}^{i-1}\end{array}\right) =
\left( \begin{array}{c} 2B_{i}-x_{n}^{i-1}\\\\-p_{n}^{i-1} \end{array}\right). \\
\end{equation}
For $E_n > V_{0}$ (refractive boundary condition), we get
\begin{equation}
\widehat{\mathcal{R}}_i \left( \begin{array}{c} x_{n}^{i-1} \\\\ p_{n}^{i-1}\end{array}\right) =
\left( \begin{array}{c} B_{i}+\dfrac{\left( x_{n}^{i-1}-B_{i}\right)p_{n}^{i} }{p_{n}^{i-1}}\\\\
\dfrac{p_{n}^{i-1}}{\lvert p_{n}^{i-1}\rvert}\sqrt{{(p_{n}^{i-1})}^2-\dfrac{2V_{0}V_{diff}}{\lvert V_{diff} \rvert}} \end{array}\right). \\
\label{refrac}
\end{equation}
\label{bound}
\end{subequations}
In this, we have used $V_{diff}=V(x_{n}^{i-1})-V(x_{s}^{i-1})$.
Thus, the dynamics of system in Eq. \ref{sham} can be described by the
standard map with $-\infty \le x_n,p_n \le \infty$ (Eq. \ref{smap1}) subjected to 
potential barriers (Eq. \ref{bc1}). Notice that by putting $V_0=0$
in Eq. \ref{refrac}, we obtain $\widehat{\mathcal{R}}_i = \mathbf{I}$ for all $i$, where
$\mathbf{I}$ is the identity matrix of order 2. Then $\widehat{\mathcal{R}}=\mathbf{I}$ and,
as expected, Eq. \ref{smap1} reduces to standard map for $V_0=0$.
Thus, the transformation (\ref{bound}) can be viewed as deviation from standard map dynamics induced
 after each encounter of the particle with the a discontinuity of potential $V_{sq}$.

\subsection{Phase space features} 
\begin{figure}
\includegraphics*[width=3.6in]{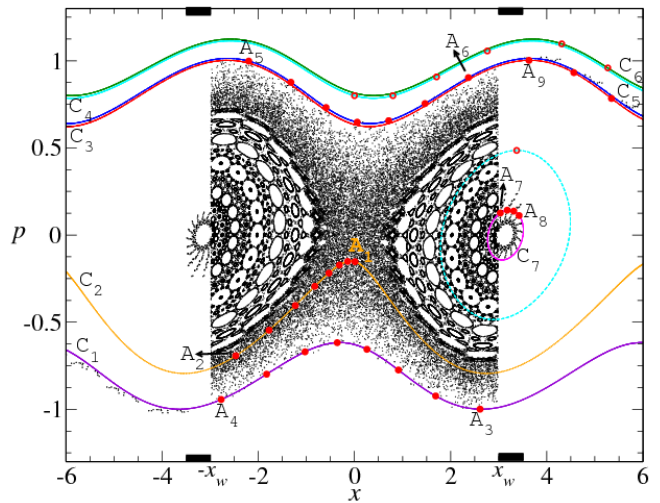}
\caption{(Color online) Stroboscopic Poincare section (black) for
$R=0.95, \epsilon=0.15, V_0=0.5, \phi=0$ and  $b=0.5$. All the continuous
curves (in color) marked $C_1$ to $C_6$ are for the corresponding standard map
with kick strength 0.15. The black box at position $x=\pm x_w$ indicates the
width $b$ of the barrier. The solid circles (in red) show a
trajectory starting from $A_1$ until it exits the potential well at $A_9$.
The time ordered sequence of the trajectory is $A_1$ to $A_2$, reflection at $-x_w$,
$A_3$ to $A_4$, reflection at $x_w$, $A_5$ to $A_6$,
cross the boundary at $x_w$, $A_7$ to $A_8$, cross the boundary at $x_w+b$, exit
the potential at $A_9$. See text for details.}
\label{psec1}
\end{figure}
Figure \ref{psec1} shows a stroboscopic section obtained by evolving the map in Eq. \ref{smap1}
for uniformly distributed initial conditions in $x \in (-x_w,x_w), p \in(-p_c,p_c)$,
where $p_c=\sqrt{2mV_0}$ is the minimum momentum required for barrier crossing.
In this paper, we have chosen kick strength $\epsilon << 1$ such that 
the corresponding standard map displays only KAM curves.
Firstly, a striking feature is the absence of invariant curves and the appearance
of a mixed phase space. This is in stark contrast with the standard map which displays
mostly quasi-periodic orbits for kick strengths of this order.
This figure also shows snap shots (solid circles in red) of trajectory in-between
successive encounters with the discontinuities at $B$.
Clearly, the evolution between two successive encounters with the
boundaries is confined to a trajectory that is identical with one of the quasiperiodic orbits
of the corresponding standard map (obtained from Eq. \ref{smap1} with $V_0=0$) shown as
continuous lines in the figure.
Due to $V_{sq}$, particle breaks away from one quasiperiodic orbit and joins
another at each encounter with the boundaries. 
This leads to absence of quasiperiodic orbits and development of mixed phase
space comprising intricate chains of islands embedded in chaotic sea.
We illustrate the effects of discontinuities in Fig. \ref{psec1} by
following a typical initial condition marked $A_1$ in the chaotic layer. This evolves
to $A_2$ on the invariant curve $C_2$ of the corresponding standard map. After a long time,
this point appears on the curve $C_1$ and goes from $A_3$ to $A_4$.
After a reflection at $-x_w$,  it goes from $A_5$ to $A_6$ on $C_3$. Then it shifts
to the barrier region $(x_w, x_w+b)$ and moves on $C_7$ from $A_7$ to $A_8$.
Depending on the winding number of the orbit in $(x_w, x_w+b)$, the particle could
have gone back in to region between the barriers or escape from the finite well.
In the present example, it makes its escape out of two barrier structures
 and its state meets the curve $C_5$ at $A_9$. Once the particle has escaped, its state evolves
on same curve as $n \to \infty$. Thus, system displays KAM behaviour for $\vert x\vert > x_w+b$.

\begin{figure}
\includegraphics*[width=2.8in]{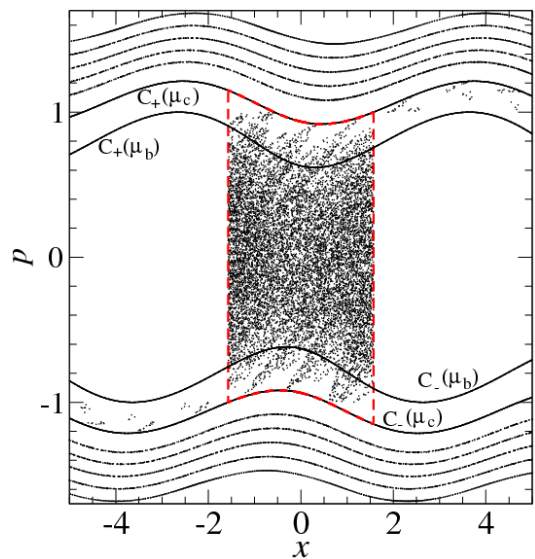}
\caption{Stroboscopic plot (excluding $C_{\pm}(\mu_b)$) for $b=10^{-3}$ and $R=0.5$.
All the other parameters are same as in Figure \ref{psec1}.
Dashed line (in red) represents the boundary of region $\cal M$.
The mild scatter of points just below $C_{+}(\mu_c)$ and just above $C_{-}(\mu_c)$ represent
the particles escaping out of the well (whose initial states were in ${\cal M}$).
The curves $C_{+}(\mu_b)$ and $C_{-}(\mu_b)$ shown here are used in section V(B).}
\label{psec2}
\end{figure}

The absence of quasiperiodic orbits can be attributed to the non-analyticity of $V_{sq}$ 
which violates the assumptions of KAM theorem. Thus, the non-KAM nature of system leads to onset
of chaos even for $\epsilon < 1$. The initial conditions starting from chaotic layer will diffuse 
in momentum space. Some of these initial conditions which reach the set of quasiperiodic
orbits $C(\mu)$ ($\mu$ being the winding number) of the corresponding standard map
which overlaps the region $|p| > p_c$ can escape from the finite well.
As $\mu$ increases, this overlap also increases and hence the escape probability is larger.
This implies that there must exist $\mu_c$ such that the states on any $C(\mu)$, with $ \mu  > \mu_c$, 
will definitely cross the barrier and escape from the well. These orbits do not encounter the
discontinuities in the potential multiple times and hence the energy of the particles evolving on 
such quasi-periodic orbits will not diffuse.
Figure \ref{psec1} also shows the trajectory of a particle (open circles in red
on the curves $C_5$ and $C_6$) in such non-diffusive region. As seen in Fig. \ref{psec1}, the 
discontinuities at $x_w$ and $x_w+b$ relocate the incoming particle from
$C_5\left(\mu_5\right)$ to another orbit $C_6\left( \mu_6\right) $, where $\mu_5$ and
$\mu_6$ are their winding numbers, respectively. As shown in Appendix B, when
$b \rightarrow 0$, the effect of these discontinuities decreases and deviation
between between two orbits measured as $(\mu_6 - \mu_5)\rightarrow 0$.
This results in the appearance of regular orbits (see Fig. \ref{psec2})
identical to those of the standard map except that the former have imperceptible
discontinuities wherever there is a discontinuity in potential. In other words, 
refraction becomes identity operation as $b \rightarrow 0$. Thus, the system shows
regular dynamics outside region enclosed between curves $C_{\pm} \left( \mu_c \right) $
 (see Fig. \ref{psec2}) as $b \to 0$. Note that the limits on chaotic phase space in terms of $\mu$
on positive and negative sides of momentum
are identical due to assumption that $\phi =0$.
Limits on the chaotic phase space would exist even otherwise, though these would
not be identical on both sides of $p=0$. The discussions in this sub-section can be summarized as follows ;
we can define a phase space
region ${\cal M}\left(\vert x\vert < x_w+b;\vert p\left( x\right) \vert < p \left( x;
\mu_c\right) \right) $, such that system has mixed phase space inside ${\cal M}$
in general and regular dynamics outside it. Here, $p \left( x;\mu_c\right)$
is momentum of any state on the curve $C_{+}\left( \mu_c\right)$ at
position $x$. In Fig. \ref{psec2}, a close numerical approximation of the region
${\cal M}$ is highlighted by the red dashed line.

We remark that for $b\rightarrow 0$, the phase space structures
inside ${\cal M}$ are
identical to those of well map that describes the dynamics of $\delta$-kicked particle in an
infinite well \cite{sankar}. This is to be expected since the well map has only reflective
boundaries for $|p| \le \infty$. Further, the well map is hyperbolic for $R<0.5$ for any
$\epsilon > 0$. The Hamiltonian in Eq. \ref{ham} also displays
complete chaos for $R < 0.5$ inside ${\cal M}$.
This is seen in Fig. \ref{psec2} as no regular structures are visible in this region
to the accuracy of our calculations. The region defined by ${\cal M}$ is determined by
the positions of potential discontinuities $\mathbf{B}$ and $C_{\pm} (\mu_c)$.
 It can be shown that $C_{\pm} (\mu_c)$ will remain 
close to $ \pm p_c (= \pm \sqrt{2mV_0} )$ when $b\rightarrow 0$ for any $\epsilon$ for which standard map 
has mostly regular phase space. 
Thus, the extent of chaotic region will depend grossly on the
positions $\textbf{B}$ and height $V_0$ of the barriers only. This implies that it
is possible to engineer chaos in a desired region by varying these parameters.

\subsection{KAM-like behavior: Role of symmetries}
\begin{figure}
\includegraphics*[width=3.5in]{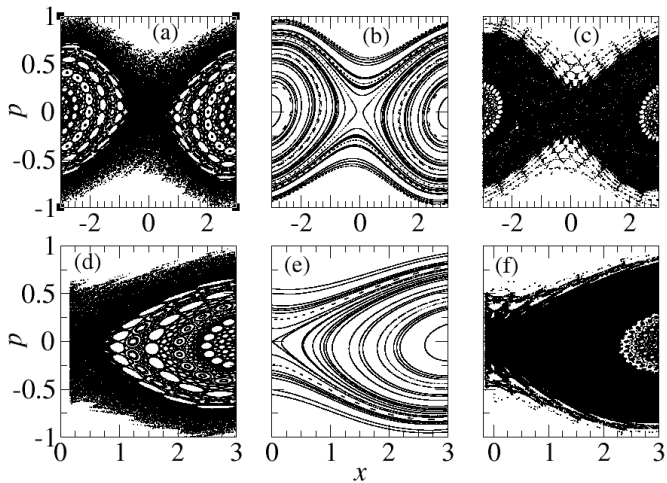}
\caption{Stroboscopic Poincare section for the Hamiltonian in Eq. \ref{ham}
showing the region $x \in (-x_l, x_r), p \in (-p_c,p_c)$
for $b=0, \epsilon=0.15, V_0=0.5$. The other parameters are (a) $R=0.95, \phi=0$
(b) $R=1.0, \phi=0$, (c) $R=1.05, \phi=0$, (d) $R=0.45, \phi=\pi/2$, (e) $R=0.5, \phi=\pi/2$
and (f) $R=0.55, \phi=\pi/2$.}
\label{psec3}
\end{figure}

In this section, we explore the conditions under which KAM or non-KAM type of dynamics
can be realized in the system. In Eq. \ref{ham}, the non-analyticity of $V_{sq}$ violates
 the assumptions of the KAM theorem. Hence, generically we expect this system to display
the signatures of non-KAM system such as the stochastic webs instead of quasi-periodic orbits
and an abrupt transition to chaos. These features are shown in Fig. \ref{psec3}(a,c,d,f).
However, we show that even in the presence of non-analyticity in $V_{sq}$,
quasi-periodic orbits similar to that in a KAM system can be realised if certain symmetry
conditions are satisfied.

As argued before, until interrupted by the barriers, the dynamics is confined
to a particular invariant curve of the corresponding standard map.
We recall that corresponding to every trajectory $C_+$ of standard map with $p_n > 0$,
there exists one and only one trajectory $C_-$ with $p_n < 0$, such that a
particle will evolve on these trajectories in exactly the same way but in opposite
direction. As shown in appendix-A, consider the $\left(R,\phi \right)$ pairs for which
the condition
\begin{equation}
\pm R\pi+\phi=l~2\pi, \;\;\;\;\; l \in \mathbb{Z}
\label{cond1}
\end{equation}
is satisfied. When Eq. \ref{cond1} is satisfied, application of $\mathcal{\widehat{R}}_i$ 
takes a
particle from $C_+$ to  $C_-$ and application of $\mathcal{\widehat{R}}_{i+1}$
brings it back to $C_+$.
This leads to quasiperiodic behavior in which the particle is confined to a pair of tori.
This quasiperiodic orbit undergoes smooth deformation,
just like in a KAM system, until it breaks for large kick strengths. Hence we call
this KAM-like behaviour for its striking resemblance to the
qualitative behaviour of a KAM system. In general, there exist infinite ($R$,$\phi$)
pairs for which KAM-like dynamical behaviour can be recovered in this system.
In Fig \ref{psec3}(b,d), we show the
sections for $R=1,\phi=0$ and $R=0.5,\phi=\pi/2$ for which KAM-like behaviour is
obtained. In Fig \ref{psec3}(a,c,d,f), we also show cases where
Eq. \ref{cond1} is not satisfied and hence for $|p|<p_c$ stochastic webs and chaotic
regions are seen.

Symmetry related invariant curves like $C_+$ and $C_-$ are due to the symmetry
of the kicking field about any $x=m\pi+\phi$ where $m$ is an integer.
It turns out that when Eq. \ref{cond1} is satisfied, kicking field is symmetric about $x_w$ 
and $x_{-w}$. 
The existence of KAM-like behaviour in presence of non-analytic potential
can be attributed to existence of centres of symmetry of kicking field at $-x_w$ and $x_w$.  

\section{Quantum dynamics}
\begin{figure}
\includegraphics*[width=3.5in]{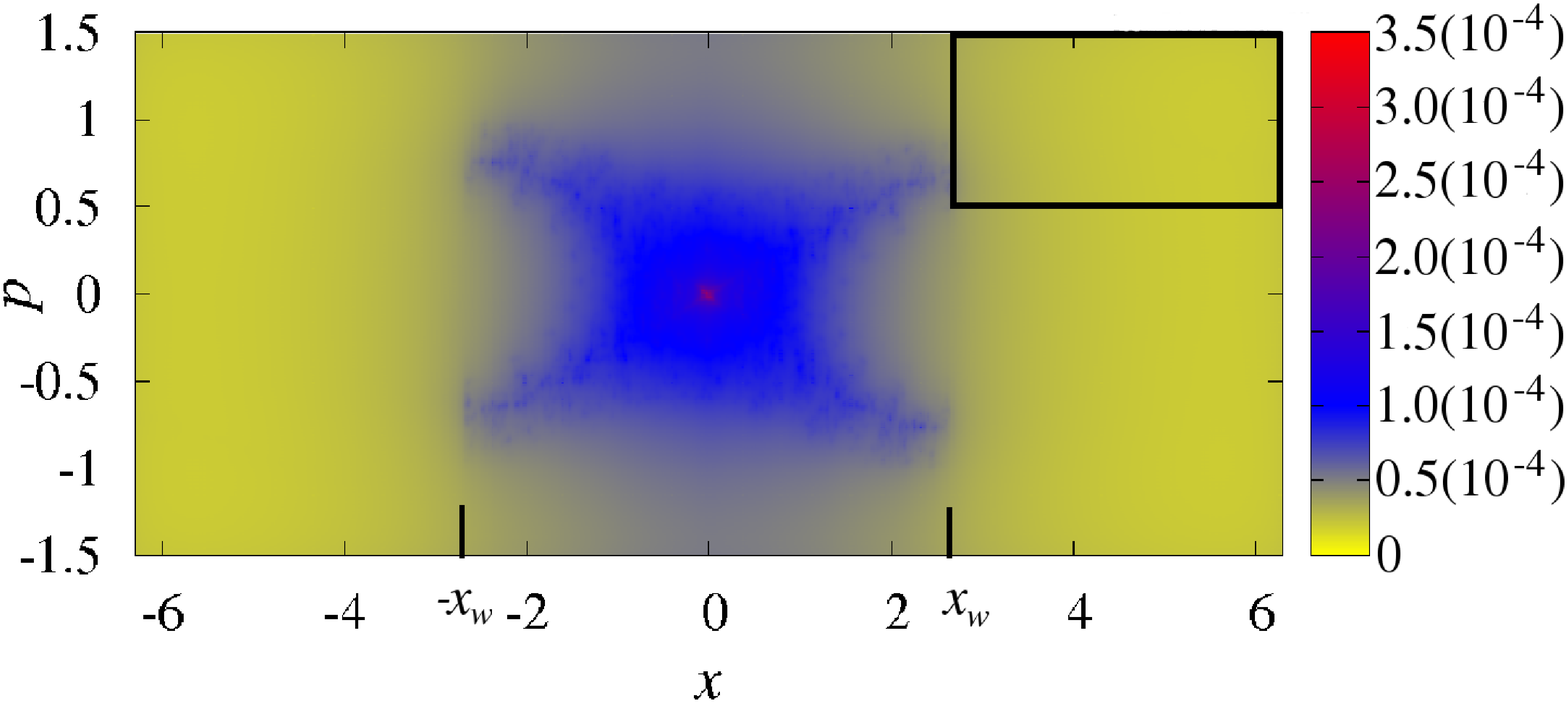}
\includegraphics*[width=3.5in]{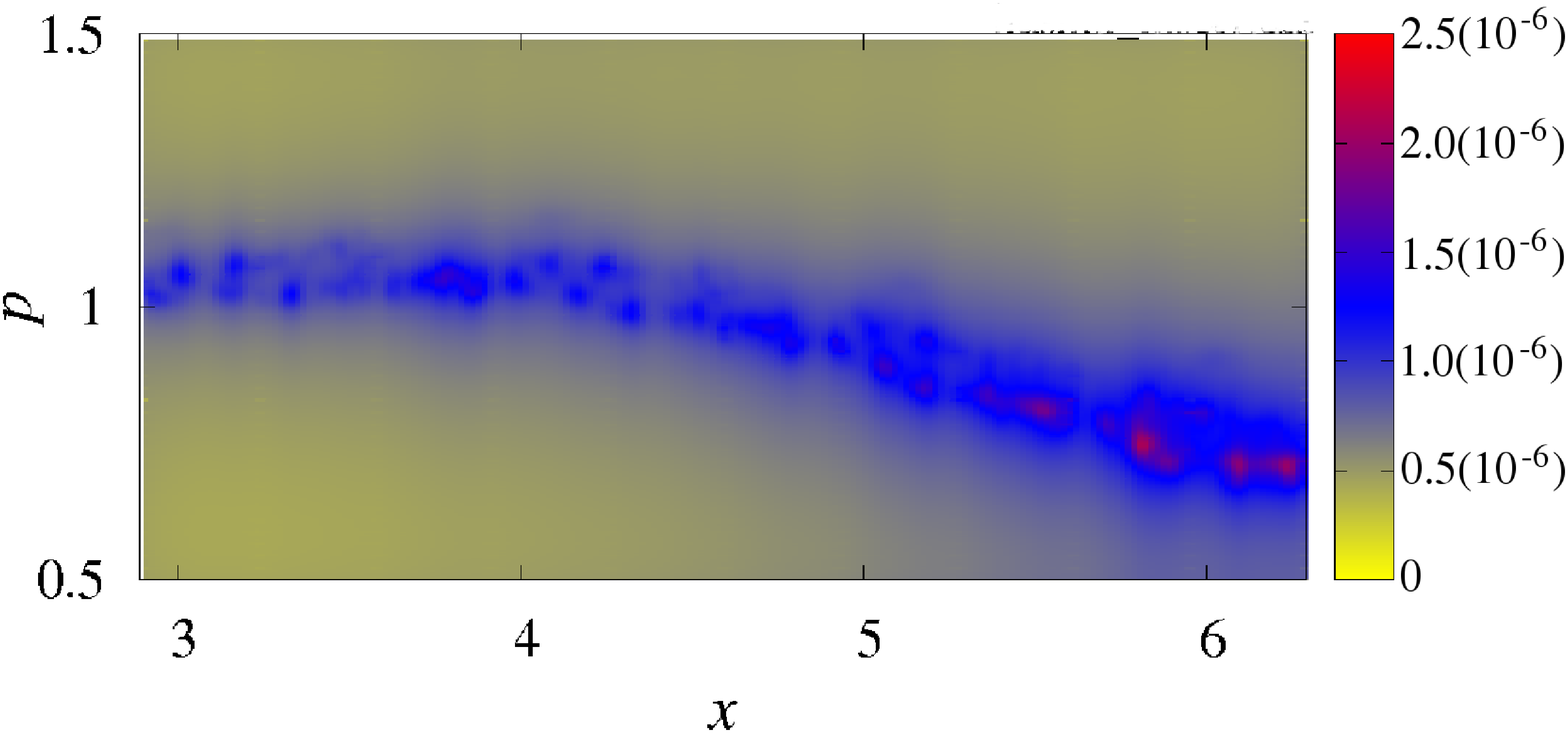}
\caption{(Color online) (Top) Husimi distribution for evolved wave packet. Initial wave function
corresponds to $Q(x_0,p_0,n) $ sharply localised inside the chaotic region around $(0,0)$.
 In the grey scale version, the regions with larger values of Husimi distribution function 
are grossly represented by the darker areas. It shows 
that the function decays very steeply outside $\left[ x_{-w} , x_{w}\right] $ and acquires 
negligible values compared to those for region inside $\left[ x_{-w} , x_{w}\right]$.
We have taken $\hbar_s = 0.0025$, $R=0.85$, $b=0.2$, $\epsilon = 0.15$, $V_0=0.5$, $\phi =0$. 
(Bottom) Enlarged  and better resolved view of inset from figure on the top shows path followed by probability
density outside the barrier region.}
\label{husimi}
\end{figure}

In this section, we discuss the quantum simulations of the system.
We start by writing down the time-dependent Schroedinger equation corresponding to
the scaled Hamiltonian in Eq. (\ref{sham}),
\begin{equation}
i \hbar_s \frac{\partial \psi}{\partial t} = 
\left[ \frac{-\hbar_s^2}{2} \frac{\partial^2}{\partial x^2} + V_{sq}
            + \epsilon \cos x \sum_n \delta(t-n) \right] \psi.
\label{tdse1}
\end{equation}
The scaled Planck's constant is $\hbar_s = \frac{2\pi^2 \hbar}{E_c T}$.
This being a kicked system, we can obtain the one-period Floquet operator,
\begin{equation}
\widehat{U} = \exp\left(-\frac{i\epsilon}{\hbar_s} \cos \widehat{x} \right)
        \exp\left( -\frac{i}{\hbar_s} \left[ \frac{{\widehat{p}}^2}{2}
  + \widehat{V}_{sq} \right] \right),
\label{floque}
\end{equation}
such that $\psi(x,n) = \hat{U}^n \psi(x,0)$. The classical limit will correspond to taking
$\hbar_s \to 0$ keeping $\epsilon = \tilde{\epsilon} \hbar_s/\hbar$ constant.
We calculate the Husimi distribution $Q(x_0,p_0,n)$ defined by
\begin{equation}
Q(x_0,p_0,n) = | \langle \psi(x,n) | x_0,p_0 \rangle |^2
\end{equation}
for a wavepacket at time $n$. In this we take $\langle x | x_0, p_0 \rangle $ as the
minimum uncertainty
wavepacket. In the semiclassical regime, the dynamics in the Husimi representation mimics the
classical dynamics of the system in phase space \cite{taka}. In Fig. \ref{husimi}, we show
the Husimi distribution at $n=250$ from which one can clearly see that the
density of Husimi distribution shows pattern similar to classical structures shown
in Fig. \ref{psec1}.

Since $\widehat{p}$ and $\widehat{V}_{sq}$ in Eq. \ref{floque} do not commute, we first divide the 
duration between successive kicks into $N_{\triangle t}$ small time steps and the
second term of Eq. (\ref{floque}) becomes $\displaystyle{\prod_{i=1}^{N_{\triangle t}}
 \exp \left( -\frac{i}{\hbar_s N_{\triangle t}} \left[ \frac{{\widehat{p}}^2}{2}  + 
\widehat{V}_{sq} \right] \right)}$. 
Then, we apply the split-operator method \cite{tannor} to evolve the system.
We use Fast Fourier transform \cite{fftw} to obtain $\widetilde{\psi}\left( p\right)$ from $\psi 
\left( x \right)$ and
vice-verse. In our calculations, we have taken $N_{\triangle t} \sim 2500$, the typical temporal 
step size
is $O(10^{-3})$ and spatial step size is $O(10^{-4})$ to ensure that the evolved
wavepackets converged to at least 8 decimal places.

The initial wavepacket at $n=0$ is located in between the two barriers.
We choose parameters $b$ and $\hbar_s$ for which the Husimi distribution
(shown in Fig. \ref{husimi}) closely resembles the classical
phase space and shows that the probability density associated with the initial
wavepacket will ultimately leave the
barrier region by predominantly following the classical path rather than by tunnelling.
Thus, the system stays in the semiclassical regime and tunnelling
is suppressed.
Quite clearly, for such a choice of parameters in the semiclassical regime, the
classical dynamical features would be reflected in the quantum dynamics as well.

In the next two sections, we discuss some interesting dynamical features, namely
(i) the non-equilibrium steady state
(ii) classically induced suppression of diffusion and (iii) momentum filtering which 
primarily arise due to co-existence of diffusive (chaotic region ($\mu< \mu_c$)) region
and non-diffusive region (regular region ($\mu > \mu_c$)) in same non-KAM system.

\section{Dynamical features}
\subsection{Non-equilibrium steady state}
\begin{figure}
\includegraphics*[angle=-90,width=3.4in]{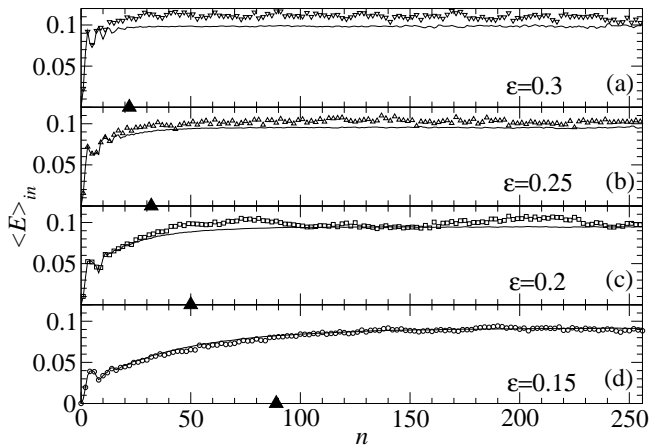}
\caption{Non-equilibrium steady state in the system in Hamiltonian Eq. \ref{ham}.
The mean energy for the particles held in between the double barrier structure. The
solid lines are the classical results and the symbols correspond to quantum results.
The other parameters are $R=0.5, b=0.2, \phi=0, V_0=0.5$ and for quantum
simulations $\hbar_s=0.0025$. The solid symbol (triangle up) marks the
time scale $t_r$ at which the system relaxes to the steady state.}
\label{nessfig}
\end{figure}

In this section, we show that the system in Eq. \ref{ham} can support non-equilibrium
steady state (NESS) for intermediate time scales. 
We start with initial conditions uniformly distributed on a thin rectangular band around $p=0$
stretched across the well region in between the potential barriers. As the 
kicking field begins to impart energy to the system, the particles which absorb
sufficient energy escape from the well. At any time $n$, the mean energy $\langle E \rangle_{in}$
of the particles lying inside the well is
$\left\langle \frac{p_n^2}{2} \right\rangle$, where $\langle . \rangle$ represents
average at time $n$ over the classical states (evolved from initial states over $n$ kicking cycles) for which 
$-x_l < x < x_r$.
In the corresponding quantum regime, we have,
\begin{equation}
\langle E \rangle_{in} = \displaystyle{\int_{-x_l}^{x_r}} ~\psi^{*}(x,n) ~
\dfrac{{\widehat{p}}^2}{2} \psi (x,n) ~dx
\end{equation}
The effect of the operator $\widehat{p}^2$ on $\psi(x,n)$ can be calculated
using fast fourier
transform and is equal to inverse fourier transform of ${p}^2 \widetilde{\psi}(p,n)$.
Figure \ref{nessfig} shows that initially $\langle E \rangle_{in}$ increases and after
a time scale $t_r$, $\langle E \rangle_{in}$ saturates to a constant.
During this time scale, the behaviour is similar to the classical diffusive regime of the
standard map.

\begin{figure}
\includegraphics*[angle=-90,width=3.5in]{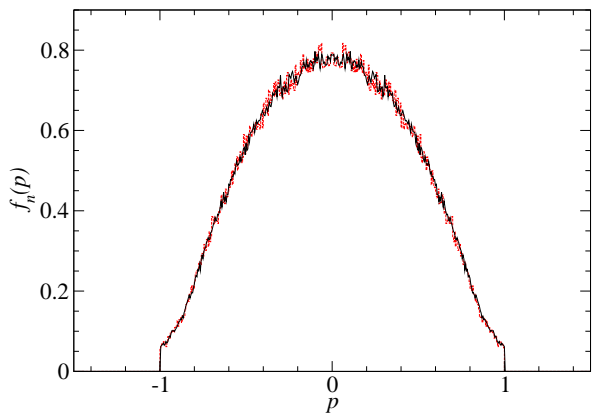}
\caption{(Color Online) Classical steady state momentum distribution for $\epsilon=0.25$
at (a) $n=100$ (solid) and (b) $n=200$ (dashed). As seen from Fig. \ref{nessfig}(b), the steady
state is reached at $n \sim 30$. The other parameters are $R=0.5, V_0=0.5, b=0.2, \phi=0$.}
\label{ness_state}
\end{figure}

The existence of steady state can be understood as follows. For the parameters used in Figure 
\ref{nessfig} the 
phase space in region $\cal M$ is fully chaotic. As kicks begin to act, any localized classical distribution 
$\rho_0(x,p)$ is
quickly dispersed throughout this region. The total energy $E_n$ 
of the particles
in the well region increases. Simultaneously, the particles with $|p| > p_c$ leave the finite well 
leading to loss of
energy. Soon the loss process becomes significant and at every kick cycle the energy lost (due to 
barrier crossings) is
more than the energy gained from the kicking potential. Thus, $E_n$ begins to decrease.
However, after the time scale $t_r$, the net energy change and the number of particles
vary in such a manner as
to maintain the mean energy $\langle E \rangle_{in}$ a constant (apart from fluctuations).
This arises because the normalised momentum distribution remains nearly invariant with time as 
shown in Fig. \ref{ness_state}.
The chaotic mixing inside the well ensures that, despite the loss of energetic particles, momentum 
distribution remain invariant.
Thus, chaos between the barriers is essential to support the NESS.
One of the factors that determine $t_r$ is the rate at which any initial distribution of 
states diffuses in the chaotic region and steady state distribution shown
in Fig. \ref{ness_state} is
achieved. This rate increases with $\epsilon$ in general. For the present case with complete chaos,
one expects this rate to be proportional to $1/\epsilon^2$, just like in the diffusive regime
of standard map and hence one expects $t_r \propto 1/\epsilon^2$. 
Numerical results shown in Fig. \ref{nessfig} show a good agreement with this gross estimate for $t_r$.

This steady state holds good until nearly all the particles have escaped out and 
only a fraction $q<<1$ remains in the well. Based on rate of diffusion in chaotic region, 
we can estimate the time at which this happens to be $t_s \propto 1/q^2 \epsilon^2$.
Since $q<<1$, we get $t_s >> 1$.
In the semiclassical regime, this mechanism
carries over to the quantum dynamics as well. Notice that $t_s$ is
larger than other relevant time scales, i.e, $t_s >> t_r > T$.  Further,
$t_s$ is typically about few hundreds of kick cycles and hence we expect
this to be experimentally accessible time scale as well.
On a much longer time scale as $t\to \infty$, all the energetic
particles escape and the steady state decays out.

Indeed, a similar non-equilibrium steady state has been experimentally observed
with periodically kicked Bose-Einstein condensate in a finite box for strong
kick strengths \cite{mark1}.
These steady states have a classical explanation. Typically, the standard kicked rotor
exhibits energy saturation and steady state, for large kick strengths, in the quantum
regime due to destructive quantum interferences. \cite{chi1,chi2}. We emphasise that 
the energy saturation, in our model as well as for the BEC in finite box \cite{mark1},
is induced by the classical effects and leaves a trail in the semiclassical regime.

Fig \ref{nessfig} shows that the quantum mean energy $\left\langle E\right\rangle_{in}$ follows the
classical curve quite closely. These results correspond to $\hbar_s=0.0025$ and reflect the 
behaviour in the semiclassical regime. Larger values of $\epsilon$ correspond
to moving away from semiclassical regime towards 
purely quantum regime. Thus, we should expect quantum averages to deviate from classical averages 
in a pronounced manner. 
This is borne out by the numerical results in Fig \ref{nessfig}(a,b,c). There is current interest 
in quantum non-equilibrium 
steady states about which not much has been explored until now \cite{marko}.
For $\epsilon >> 1.0$, the quasiperiodic orbits of the standard map are
sufficiently destroyed to allow global 
transport in phase space. Then, particles do not have to rely on discontinuities in 
$V_{sq}$ to diffuse in phase space. This leads
to unlimited energy absorption by the particles between the barriers and NESS is
not supported. Then, the system essentially works like the kicked rotor in the
strongly chaotic regime.

\subsection{Energy saturation and steady state}
As pointed out earlier, the region covered by the quasi-periodic orbits
$C(\mu)$ with $\mu > \mu_c$ is non-diffusive. Now, consider curves $C_{\pm}(\mu_b)$
such that maximum value of $\vert p \vert$ for both these curves is equal to $p_c$ as shown in
Fig. \ref{psec2}. Then, all the curves $C(\mu)$ with $\mu < \mu_b$ will have $p \in
\left[ -p_c, p_c\right]$. So any state evolving on
one of them will always get reflected at the barriers. Thus, to escape from
the finite well, every phase space point in the chaotic region
must first reach any $C(\mu)$ with $\mu_b <\mu < \mu_c$.
As time $n \to \infty$, all the particles would have escaped from the well
and get locked on to one of the invariant curves $C(\mu)$ of the
corresponding standard map. Thus, the momenta of escaping particles settle
to a stationary distribution on $C(\mu)$ with $\mu_b <\mu < \mu_c$.
Thus, the momentum distribution reaches a steady state as $n \to \infty$
and their mean energy
$\left\langle E \right\rangle$ saturates to $\left\langle E \right\rangle_s$.

\begin{figure}
\includegraphics*[angle=-90,width=3.5in]{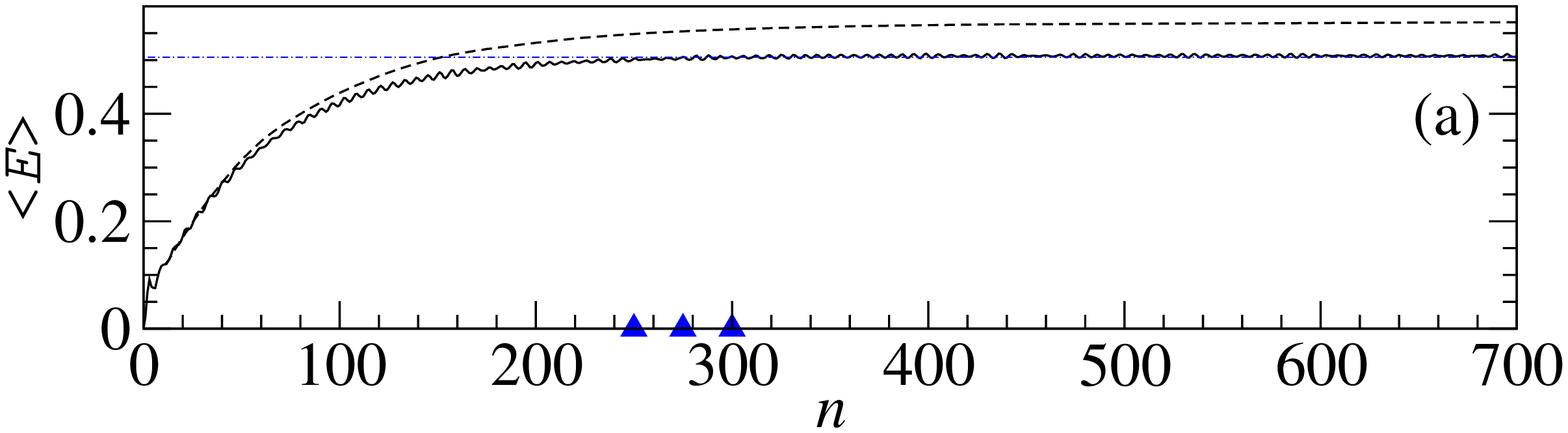}
\includegraphics*[angle=-90,width=3.5in]{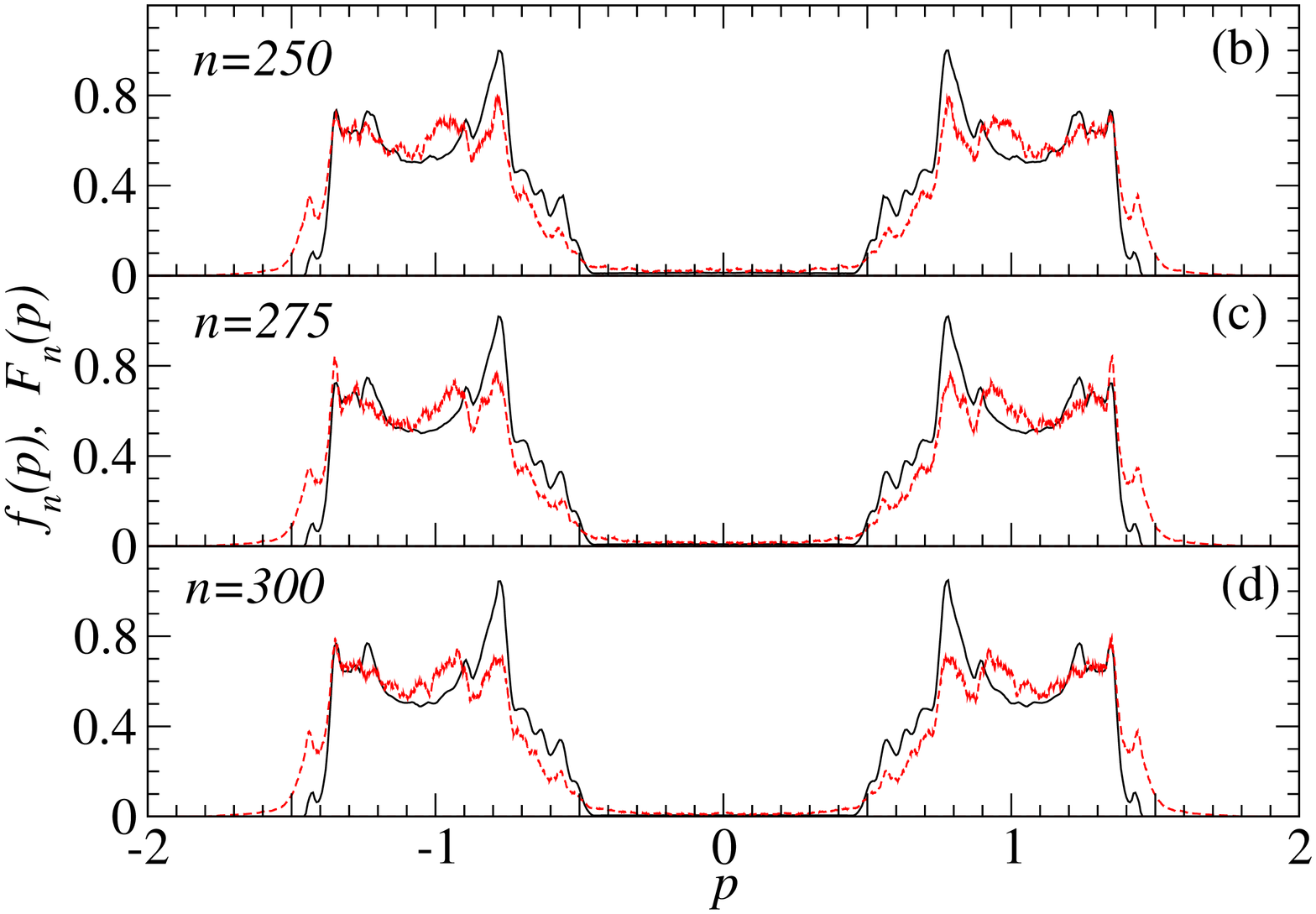}
\caption{(Color Online) (a) Classical (solid line) and quantum (dashed line) mean
energies as a function of time $n$ for $\epsilon =0.3$. Other parameters are same as 
in Fig. \ref{nessfig}. Numerically estimated value of $\left\langle E 
\right\rangle_s$
for classical system is shown as a broken line. The triangles in the $x$-axis are the
times for which momentum distribution are shown in (b,c,d).
Classical (solid line) and quantum (dashed line) momentum distributions
at (b) $n=250$, (c) $n=275$ and (d) $n=300$. Note that the distributions are nearly identical.}
\label{ness-local}
\end{figure}

In Fig. \ref{ness-local}(a), the broken curve (blue) shows $\left\langle E\right\rangle_s$
for the classical system.
As this figure shows, the mean energy of the system increases with time and asymptotically
approaches $\left\langle E\right\rangle_s$. In the semiclassical regime, we expect
a similar behaviour for the quantum average and this is shown as dashed curve in Fig. \ref{ness-local}(a).
The small difference in saturated values of quantum and
classical mean energies can be attributed to the finiteness of Planck's constant which makes
its effect felt as $\epsilon$ increases.

Further, Fig. \ref{ness-local} (b,c,d) also shows the classical momentum distribution $f_n(p)$
and its quantum analogue $F_n(p) = |\widetilde{\psi}(p,n)|^2$
for the same set of parameters after evolving the system for $n=250$, 275 and 300 kicking periods.
Probability distribution in position representation (not shown here) reveals that at
$n=250, 275$ and 300 the probability density in between the barriers is negligible.
Nearly identical distributions in Fig. \ref{ness-local}(b,c,d) mark the existence of steady state.
Notice that small departures from semiclassical
regime is also visible here in the form of slight difference between classical and quantum 
distributions.
For the energy saturation effect, complete chaos between barriers is not essential. If some
sticky islands are present between the barriers, the saturated
classical and quantum distributions as $n \to \infty$ will display a non-zero component in the 
finite well region.
These non-chaotic components tend to remain localized and will never escape out.

\subsection{Momentum filtering}

\begin{figure}
\includegraphics*[angle=-90,width=3.3in]{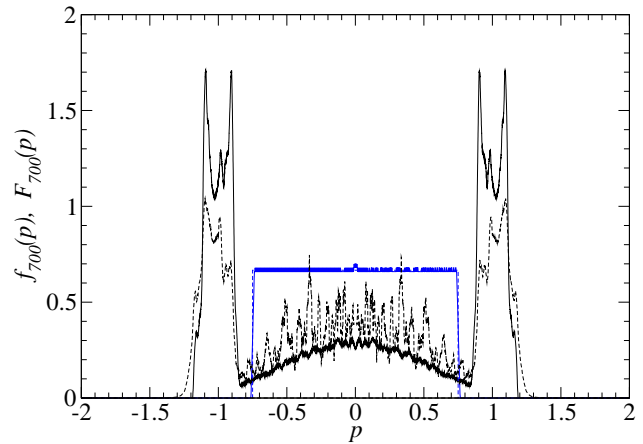}
\caption{(Color Online) (a) Classical (solid line) and quantum (dashed line)
momentum distributions at $n=700$ are displayed for $V_0 = 0.5$, $b=0.2$, $R =0.5$, $\epsilon =0.1$. For quantum
simulation, $\hbar_s = 0.0025$. The initial 
distribution at time $n=0$ is 
a uniform distribution, the rectangular curve shown in blue.
See text for details.}
\label{mfilt}
\end{figure}

As demonstrated in section 5(B), when all the chaotic particles exit from the finite well region,
a steady state is reached. One possible manifestation of this asymptotic state is the momentum
filtering effect that occurs for certain choices of parameters. It is possible to choose
system parameters such that momentum distribution of escaped particles becomes narrow.
Thus, any broad initial momentum distribution at $n=0$, after sufficient kicking periods,
leads to a distinctly narrow momentum distribution. This is shown in Fig. \ref{mfilt}.
In this figure, the initial conditions are uniformly distributed in the chaotic layer lying in 
between the barriers.
This chaotic layer also ensures that the final result is independent of the details of the
initial distribution. The figure shows the momentum distributions $f_{700}(p)$ (classical) and 
$F_{700}(p)$ (quantum)
plotted for $n=700$. By this time, a large fraction
of particles have escaped from the well and the distribution has become bimodal with distinct peaks
near $-p_c$ and $p_c$. This shows that the double barrier
structure, in presence of the kicking field, acts as a momentum filter.
We obtain filtering effect for a range of kick-strengths (not shown here) and observe that with
decrease in $\epsilon$, the two bands in bimodal distribution become narrow. However, the time at
which system approaches steady state corresponding to this bimodal distribution becomes very large.
Indeed, since it is experimentally possible to design barrier heights of desired choice,
it will be be possible to use double barrier structure to produce filter with desired
value of $p_c$.
From Fig. \ref{psec1}, we note that the escaped particles follow extremely close set of invariant
curves and their speed, averaged over time, will converge to the winding number of the orbits
involved. Hence, the speed distribution will have peaks of infinitesimal width at $|p_c|$.
It is pertinent to note that a momentum
filtering effect based on a very different
mechanism has been studied by Monteiro et. al. in the context of a variant
of kicked rotor model \cite{mont1}.

We emphasise that all the dynamical features discussed in section 5(A,B,C) can be explained on 
the basis of
(i) co-existence of diffusive and non-diffusive regions which exists because the non-KAM nature
of the system affects the dynamics differently in different phase space regions, and
(ii) presence of KAM curves through out the phase space outside the double barrier region.
Hence, all the dynamical features
can be attributed to the interplay between the KAM and non-KAM behaviour of the system.

\section{Discussions and summary}
In summary, we have presented primarily numerical results of the dynamics
of non-interacting particles in a double barrier structure acted upon by
periodic kicking field. This model differs from the paradigmatic kicked rotor.
This is essentially a non-KAM system and hence chaotic dynamics sets in
for even for infinitesimal excursions from the integrable limit of kicking
strength $\epsilon=0$. Further, this displays non-equilibrium steady state and
classically induced suppression
of energy growth in the semiclassical regime. This is in contrast with the
classical kicked rotor that displays diffusion only for $\epsilon >> 1$ and its quantum version
arrests this through dynamical localisation, an outcome of quantum
interferences.

Some of the earlier works on the double barrier type potential have 
considered it as a scattering problem, in a different setting with a drive term.
For example, see references \cite{dbar}.
An incoming wavepacket hits
the left barrier (see Fig. \ref{scheme}) and tunnels in to it and,
depending upon the parameters
chosen, some or all of it emerges out of the right barrier. This mechanism requires
purely quantum effects such as tunneling and in this work we have
deliberately avoided them to focus on the semiclassical regime.
Since tunneling probability is nearly zero in this semiclassical
setting, any initial distribution
placed anywhere outside the barriers $(|x| > a+b)$ will continue to evolve
on the KAM like invariant tori.
However, based on the results obtained in this paper, we can speculate
about the case when quantum effects come into play. Tunneling will allow
a wavepacket to enter the through the left barrier and non-KAM chaos
will ensure that it gets dispersed. But now, the wavepacket can tunnel
out through the right barrier. This scenario could potentially lead to an
interesting competition between above barrier crossings and tunneling.
Another interesting case relates to periodic version of this model 
which can also be used for directed transport. We are pursuing these
questions and will be reported elsewhere.

 The dynamical features in our model such as the non-equilibrium steady state and
classically induced energy growth suppression are of current interest
in the general context of transport and localisation especially for
interacting systems such as the Bose-Einstein condensates. Recently there have
been several experimental results that point to classical features suppressing
energy growth of condensates \cite{bec-trans}. Typically, in such experiments,
condensates are released from
a confining potential and their expansion in a disordered potential is studied.
When chemical potential $\mu < V_0$, where $V_0$ is the strength of disorder,
condensates are classically reflected from the fluctuations of the disordered
potential effectively localising the condensates. In our model, particles
are neither interacting nor there is any disordered potential. However,
the non-KAM chaotic dynamics and KAM like invariant curves provide the
essential ingredient for the suppression of diffusion. Even as the particles are
transported in the position space their energy absorption is restricted
as $t \to \infty$ by KAM like structures.
Such studies form an important background
to understand and clearly distinguish similar quantum phenomena like the Anderson
localisation from the classically induced ones and also to explore the
connections between interactions, localisation and disorder.

 Quantum chaos in double barrier potentials have been studied before
experimentally using GaAs/AlGaAs heterostructures \cite{dbleb} though not with a
periodic kicking field. In these experiments electrons tunnel through the
double barrier potential and chaos is induced within the barriers due to the
field created by the charge accumulation in the well \cite{dbleb}. Since resonant
tunnelling plays an important role in this experiment, this
can be regarded as being quantum in nature without classical
analog. The double barrier system in Eq. \ref{ham} could be used
with resonant tunnelling to study purely quantum effects as well
though in the present work we have primarily explored the classical and
semiclassical features. The foregoing arguments also imply that the
system can also be realized experimentally in a laboratory. The cold atoms
in optical lattices is the testing ground for variants of kicked rotor.
An experimental set-up involving cold atoms, optical lattices with
double barrier heterostructures should be possible.

Currently there is considerable interest in the exciting field of
chaotic ratchets \cite{rat}. Generally, ratchets are systems with broken spatio-temporal symmetries
from which directed transport can be obtained even in the absence of a net bias.
There have been several proposals and at least one
experimental realization for a chaotic ratchet in the last few years.
The system presented in this work lacks the spatial periodicity required of a ratchet.
But the kicking potential, being sinusoidal, is already spatially periodic.
Further, from a theoretical perspective, it is not difficult to have spatially periodic double 
barrier structures.
Then, it might become possible to realize ratchet dynamics in this system. All the
existing chaotic ratchet proposals are based on systems that obey KAM theorem.
The model presented in this work might lead to new ways to use non-KAM type dynamics
for deterministic, directed transport.

\acknowledgments 
The authors acknowledge many useful discussions with Dilip Angom
during the course of this work. Numerical calculations for quantum system 
are carried out on PRL 3TFLOP cluster computer. One of the authors (HP) 
thanks Manjunatha of CDAC and computer center staff for suggestions 
and assistance in efficiently using cluster computer.

\renewcommand{\theequation}{A\arabic{equation}}
\setcounter{equation}{0}   
\section*{APPENDIX A}

Consider a particle that evolves on an invariant curve of the standard map
$C_5(\mu_5)$, approaches right barrier at $x_w=R\pi$ with $p>p_c$ during its motion
after $n^{th}-$kick, crosses it and exits on to another invariant curve of
standard map $C_6(\mu_6)$. In this appendix, we show that as the width
of the barrier $b \to 0$, $C_5(\mu_5) \to C_6(\mu_6)$.

After the particle crosses the interface at $x_w$ and if $\Delta t$ denotes the time
it will take to cross the barrier region of width $b$, then $\Delta t\to 0$ if $b\rightarrow 0$.
Hence, the probability that a particle will experience the next 
kick while crossing the barrier will also tend to zero. Hence we can assume
that the particle does not experience a kick while crossing the barrier.
In such a situation, the particle will face only two discontinuities between 
$n$th and $(n+1)$th kick. Thus, $k=2$, $B_1=x_w$ and $B_2=x_w+b$.
From our assumptions, $\left( \begin{array}{c} x_n^{0}\\p_n^{0} \end{array}\right)$
lie on $C_5\left( \mu_5\right) $, and $\left( \begin{array}{c} x_n^{2}\\p_n^{2} \end{array}\right)$
will lie on $C_6\left( \mu_6\right)$. 

\begin{equation}
\left( \begin{array}{c} x_n^{1}\\p_n^{1} \end{array}\right)
= \widehat{\mathcal R}_1\left( \begin{array}{c} x_n^{0}\\p_n^{0} \end{array}\right)  \Rightarrow
\left( \begin{array}{c} x_w + \dfrac{\left( x_n^{0}-x_w\right) p_n^1}{p_n^0}
\\\sqrt{{\left( p_n^{0}\right) }^2-2V_0} \end{array}\right)
\label{eqa1}
\end{equation}
Similarly,
\begin{equation}
\left( \begin{array}{c} x_n^{2}\\p_n^{2} \end{array}\right)
 = \widehat{\mathcal R}_2\left( \begin{array}{c} x_n^{1}\\p_n^{1} \end{array}\right) \Rightarrow
\left( \begin{array}{c} x_w + b + \dfrac{\left( x_n^{1}-x_w-b\right) p_n^2}{p_n^1}
\\\sqrt{{p_n^{1}}^2-2V_0} \end{array}\right).
\label{eqa2}
\end{equation}
Substituting for $x_1$ and $p_1$ from Eq. \ref{eqa1} in Eq. \ref{eqa2}, we get,
\begin{equation}
\left( \begin{array}{c} x_n^{2}\\p_n^{2} \end{array}\right)
= \left( \begin{array}{c} b- \dfrac{bp_n^0}{p_n^1} + x_n^0 \\\ p_n^0 \end{array}\right)
\end{equation}
Using $b\rightarrow0$, we get, $\left( \begin{array}{c} x_n^{2}\\p_n^{2} \end{array}\right)
\rightarrow \left( \begin{array}{c} x_n^{0}\\p_n^{0} \end{array}\right)$. This implies
$C_5(\mu_5) \rightarrow C_6(\mu_6)$ or $\mu_6-\mu_5 \rightarrow 0$.

\renewcommand{\theequation}{B\arabic{equation}}
\setcounter{equation}{0}   
\section*{APPENDIX B}

We show that for certain special choices of $(R,\phi)$, reflection from the walls of potential
$V_{sq}$ takes a state from invariant curve $C_+$ to its symmetric counterpart $C_-$, where $C_+$ 
and $C_-$ are
related through reflection symmetry about $\left(0,0 \right) $. Let 
\begin{equation}
\left\lbrace\begin{array}{c}R\pi+\phi=l\pi \\-R\pi+\phi=m\pi \end{array}\right\rbrace,
\;\; l,m \in \mbox{integer}
\label{eqb1}
\end{equation}
Then, $x_r=l\pi$ and $-x_l=m\pi$.
Let $\left( \begin{array}{c} x_n^{i-1}\\p_n^{i-1} \end{array}\right)$ lie on $C_+$.
Reflection from the right boundary at $x_r$ will take it to
\begin{equation}
\left( \begin{array}{c} x_n^{i}\\p_n^{i} \end{array}\right)=
\widehat{\mathcal R}_i\left( \begin{array}{c} x_n^{i-1}\\p_n^{i-1} \end{array}\right)=
\left( \begin{array}{c} 2l\pi-x_n^{i-1}\\-p_n^{i-1} \end{array}\right)
\end{equation}
on the invariant curve $C$.
The spatial periodicity of $2\pi$ in the standard map implies that
\begin{equation}
\left( \begin{array}{c} \left( 2l\pi-x_n^{i-1}\right) \mbox{mod} \left( 2\pi\right)\\-p_n^{i-1} 
\end{array}\right)
         =  \left( \begin{array}{c} -x_n^{i-1}\\-p_n^{i-1} \end{array}\right)
\end{equation}
is on $C$. Since $\left( \begin{array}{c} -x_n^{i-1}\\-p_n^{i-1} \end{array}\right)$ 
is on $C_-$ and $C_-$ is unique, we have $C=C_-$.
Thus, the effect of reflection from the right boundary at $x_r$ is to take a
state from $C_+$ to $C_-$ 
if Eq. \ref{eqb1} is satisfied. Similarly, the effect of reflection from left boundary
at $-x_l$ is to take a state from $C_-$ to $C_+$.

\end{document}